\documentclass[aps,prl,twocolumn,showpacs]{revtex4}
\usepackage[dvips]{graphicx}
\usepackage{times}
\begin{document}
\draft

\title{Critical Current Oscillations in Strong Ferromagnetic Pi-Junctions}
\author{J. W. A. Robinson$^1$, S. Piano$^{1,2}$, G. Burnell$^1$, C. Bell$^3$, and M. G. Blamire$^1$}
\address{1. Department of Material Science, University of Cambridge, Pembroke Street, Cambridge CB2 3QZ,
UK\\
2. Physics Department, CNR-Supermat Laboratory, University of Salerno, Via S. Allende, 84081 Baronissi (SA), Italy\\ 3. Kamerlingh
Onnes Laboratory, Leiden University, P.O. Box 9504, 2300 RA Leiden,
The Netherlands}

\begin{abstract}

We report magnetic and electrical measurements of Nb Josephson
junctions with strongly ferromagnetic barriers of Co, Ni and
Ni$_{80}$Fe$_{20}$ (Py). All these materials show multiple
oscillations of critical current with barrier thickness implying
repeated $0$-$\pi$ phase-transitions in the superconducting order
parameter. We show in particular that the Co barrier devices can be
accurately modelled using existing clean limit theories and so that,
despite the high exchange energy (309 meV), the large I$_c$R$_N$
value in the $\pi$-state means Co barriers are ideally suited to the
practical development of superconducting $\pi$-shift devices.

\end{abstract}

\date{21 July 2006}

\maketitle

Although the interplay of superconductivity and ferromagnetism has
been the subject of study for many decades \cite{Buzdin05},
theoretical and experimental investigations into the properties of
superconductor-ferromagnetic metal (S/F) heterostructures have seen
an upsurge in interest in recent years following the experimental
observation of $0$ to $\pi$ transitions in the superconducting order
parameter in S/F thin films by Ryanzanov \textit{et al.}
\cite{RyazanovMar01} and by Kontos \textit{et al.} \cite{Kontos02}.
In terms of the Josephson relationship $I_c = I_0 \sin \Delta\phi$,
where $\Delta \phi$ is the phase difference between the two S
layers, a transition from the $0$ to $\pi$ states implies a change
in sign of $I_0$ from positive to negative. Physically, such a
change in sign of $I_0$ is a consequence of a phase change in the
electron pair wave function induced in the F layer by the proximity
effect. Experimentally, measurements of $I_c$ are insensitive to the
sign of $I_0$ and hence the absolute $I_c$ is measured; this implies
that a change in state from $0$ to $\pi$ will lead to a zero
crossing of $I_c$ and a sharp cusp at $I_c = 0$. It is possible to
describe the $I_c$ dependence with ferromagnetic thickness ($d_F$)
by the generic expression
\begin{equation}
\ I_c R_N (d_F)\propto  I_c R_N (d_0)\Bigg | \frac{\sin
\frac{d_F-d_1}{\xi_{2}}}{\sin \frac{d_F-d_0}{\xi_{1}}} \Bigg| \exp
\bigg\{  \frac{d_0-d_F}{\xi_{1}} \bigg\}, \label{cleanlimit1}
\end{equation}
where $d_1$ is the thickness of the ferromagnet corresponding to the
first minimum and $I_c R_N (d_0)$ is the first experimental value
of $I_c R_N$ ($R_N$ is the normal state resistance). Transitions can
be observed as oscillations in the critical temperature ($T_c$) of
S/F multilayers \cite{aJiang,bhge,cObi,dLazar} as well as
oscillations in the $I_c$ of S/F/S junctions with both thickness of
the F layer \cite{cbellprb05,Blum,Blum2006} and, for weak
ferromagnets whose exchange energy ($E_{ex}$) is comparable to $k_B
T_c$ of the superconductor, with temperature \cite{sellier}.

The majority of studies of S/F/S structures have concentrated on the
use of weak ferromagnets, such as Cu$_x$Ni$_{1-x}$ and
Pd$_x$Ni$_{1-x}$, where the temperature dependence can be observed
and where thickness of the ferromagnetic layer over which
oscillations in critical temperature or current are observed can be
comparatively large. Even where strong ferromagnets have been used,
a significant magnetic `dead' layer corresponding to a loss in total
moment \cite{Pick} is usually observed which complicates the
modeling - see table \ref{table}.

In the dirty limit where the mean free path $L<d_F$ and $L< \hbar
v_f / E_{ex}$ the two decay lengths, $\xi_1$ and $\xi_2$, take
similar values and so multiple oscillations of $I_c$ are not
observed. In contrast, in the clean limit where $d_F<L$ and $L>
\hbar v_f / E_{ex}$ the decay of the envelope determining the
modulation amplitude ($\xi_1$) can be much larger than the
oscillation period ($\xi_2$). Most previous studies, including those
using strong ferromagnets, have been performed in the dirty limit;
for practical applications, in which large $I_c R_N$ values are
required in the $\pi$ state it is vital to develop high quality
clean limit S/F/S devices. A recent report using the ferromagnetic
intermetallic Ni$_3$Al shows $I_c$ oscillations in the clean limit
\cite{Born2006}, but with a very large magnetic dead layer which is
not accounted for in any phenomenological model and which is likely
to make practical control of the phase state of the junction
difficult.

Co is a proven device material which can be deposited in clean thin
film form with accurately controlled thickness; however, it has not
been previously applied in S/F/S junctions because the exchange
energy was considered to be far too large. In this letter, we report
for the first time measurements of junctions containing Co barriers,
together with comparative studies of Py and Ni barriers. We show
that, unlike Py and Ni and the weak ferromagnets, the Co data fits
excellently to clean limit theory. As importantly, the magnetic dead
layer in the Co is less then $1$ nm allowing precise control of the
phase state of Nb/Co/Nb $\pi$-junctions.


Nb/Co/Nb, Nb/Py/Nb and Nb/Ni/Nb films were deposited on $10\times 5$
mm silicon (100) substrates coated with a 250 nm thermal oxide.
Simultaneously to growing $10\times 5$ mm thin films for patterning,
identical $5\times 5$ mm films were deposited for magnetic
characterization in a vibrating sample magnetometer (VSM). All of
the layers were deposited by d.c. magnetron sputtering at 1.5 Pa and
the deposition system was baked out for seven hours and subsequently
cooled with liquid nitrogen for at least one hour prior to the
deposition, which gave a base pressure better than $5\times 10^{-6}$
Pa and an oxygen partial pressure of less than $3\times 10^{-9}$ Pa.
The deposition rates are: $ 1.2 nm min^{-1}$ for Co, $ 1.6 nm
min^{-1}$ for Py, $ 0.4 nm min^{-1}$ for Ni, $ 2.4 nm min^{-1}$ for
Cu and $ 12.6 nm min^{-1}$ for Nb.
 In a single deposition run, multiple silicon substrates were placed on a
rotating holder which passed in turn under three magnetrons. The
thickness of each layer was controlled by setting the angular speed
at which the substrates moved under the respective targets and by
setting the target power. When depositing Co, Py, and Ni barriers,
an acceleration curve was programmed which allowed the angular speed
of the substrates to change monotonically as the substrates passed
under the relevant targets. The thickness of the Co ($d_{Co}$), Py
($d_{Py}$) and Ni ($d_{Ni}$) was then dependent on the substrate
position, $\theta$, on the rotating holder. With knowledge of the
deposition parameters, the rotation was programmed such that
$d(d_{F})/d\theta$ was a constant. This method of varying $d_F$
guaranteed in all cases that the interfaces between each layer
prepared under the same conditions and the only variation was the
thickness. To confirm control over the thicknesses we performed
X-ray reflectivity of Nb/Co/Nb thin films where the Nb layers had a
thickness of $5$ nm and the Co barrier thickness was varied from
$0.5$ nm to $5.0$ nm. A series of low angle X-ray scans were made
and the thickness of the Co layer ($d_{Co(observed)}$) extracted by
fitting the period of the Kiessig fringes using a simulation
package. It was found that our expected thicknesses,
$d_{Co(expected)}$, was well correlated with $d_{Co(observed)}$ with
a mean deviation of $0.2$ nm.

\begin{figure}[t!]
\centering
\includegraphics[width=7cm]{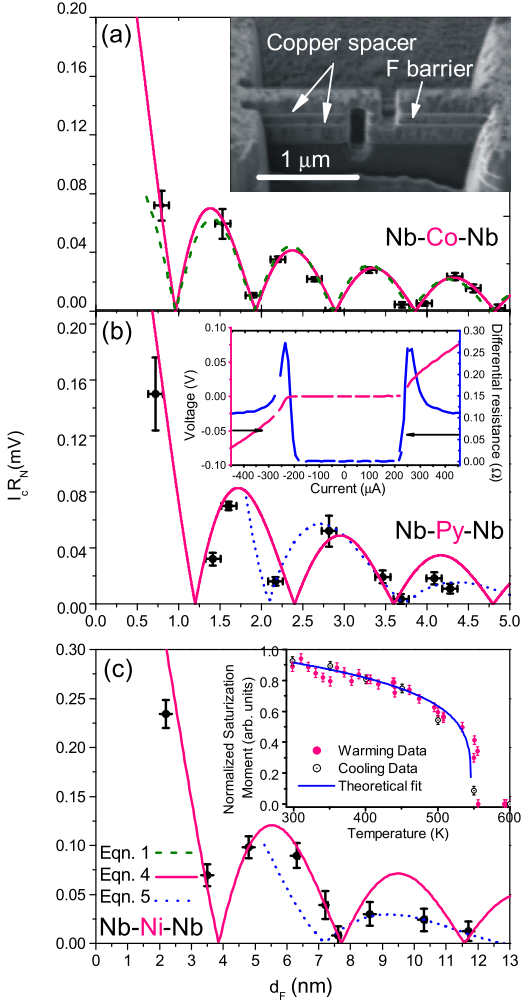}
\caption{(Color online) Characteristic voltage as a function of Co,
Py and Ni thickness at 4.2K. The solid lines are fits to Eq.
(\ref{cleanlimit}), the dotted lines are fits to Eq.
(\ref{dirtylimit}), and the dashed line is a fit to Eq.
(\ref{cleanlimit1}) as described in text. Inset (a): An FIB
micrograph of a typical Nb-Co-Nb Josephson junction. Inset (b):
$V(I)$ and $dV/dI(I)$ plotted for a Nb/Py/Nb Josephson junction at
4.2K. Inset (c): normalised magnetization as a function of warming
and cooling temperature for a Nb/Ni/Nb trilayer where $d_{Ni}\simeq
9$ nm. \label{merge}}
\end{figure}

To assist in locating the barrier layer in subsequent focused ion
beam (FIB) processing, a thin ($20$ nm) normal metal layer of Cu was
embedded in the $250$ nm thick Nb electrodes located $50$ nm from
the ferromagnetic layer (a distance greater then the coherence
length of Nb). At $20$ nm, the Cu layer is much thinner than the
coherence length in Cu and is therefore totally proximitised by the
Nb and plays no part in the electrical properties of the junctions.
The thicknesses of the F barriers for the Co and Py junctions was
varied from approximately $0.5$ nm to $5$ nm and the Ni barrier
thickness was varied from $1.0$ nm to $10$ nm. The films were
patterned using optical lithography, followed by broad beam Ar ion
milling ($3$ mAcm$^{-2}$, $500$ V beam) to produce micron scale
tracks and contact pads, to allow four-point measurements to be
performed. These tracks were then patterned with a Ga$^+$ FIB
(Philips/FEI FIB 200) to achieve vertical transport with a device
area in the $0.2-1$ $\mu$m$^2$ range. The FIB technique for
processing such junctions is described in detail elsewhere
\cite{BellNano03}. A micrograph of one of the junctions processed
using the FIB for this work is shown as the inset of Fig.
\ref{merge}(a).

\begin{figure}[t!]
\centering
\includegraphics[width=8cm]{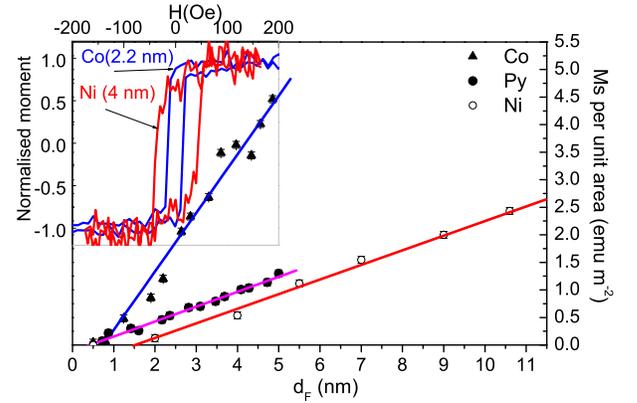}
\caption{(Color online) Scaling of the magnetic moment per unit area
vs. Co, Py and Ni barrier thickness at room temperature. Inset:
hysteresis loops for Co and Ni at room temperature.
\label{deadlayer}}
\end{figure}

To investigate the magnetic properties of our films we have
measured, using a VSM at room temperature, the magnetic moment per
unit surface area of the films as a function of $d_F$ (see Fig.
\ref{deadlayer}). Extrapolating the least-squares fit of the data in
Fig. \ref{deadlayer} gives magnetic dead layers of $\simeq 0.75$ nm
for Co, $\simeq 0.5$ nm for Py and $\simeq 1.5$ nm for Ni. The
causes of a magnetic dead layer are attributed to factors such as
lattice mismatch causing elastic deformation, formation of amorphous
interfaces and a breakdown in the crystal structure leading to a
reduced exchange interaction between neighbouring atoms and hence a
reduction in $T_{Curie}$ and $E_{ex}$ \cite{JAarts, Renjun,Qunwen}.
In the case of Co and Ni, both the thickness of the dead layer and
the total moments for a given thickness greater than the dead layer
are close to those reported in systematic studies Nb/F bilayers
\cite{cObi,mattson1997}.

Transport measurements were made in a liquid He dip probe. The
differential resistance as a function of bias current of the
junctions was measured with a standard lock-in technique. The $I_c$
was found from the differential resistance as the point where the
differential resistance increases above the value for zero bias
current. $R_N$ was measured using a quasi-dc bias current of 3-5 mA,
this enabled the nonlinear portion of the I-V curves to be
neglected, but was not large enough to drive the Nb electrodes into
a normal state. A $dV/dI(I)$ and $V(I)$ plot for a typical Nb-Py-Nb
Josephson junction is shown inset of Fig. \ref{merge}(b). In
general, the critical currents of the devices measured for this work
ranged from $500$ $\mu$A to below the minimum sensitivity of our
apparatus ($50$ nA), while the normal state resistances were in the
range $1.0$ m$\Omega$ to $100$ m$\Omega$.


The variation of $I_c R_N$ as a function of Co, Py and Ni thickness
at 4.2 K is shown in Figs. \ref{merge}(a-c) respectively. Each point
in these Figs. corresponds to the mean of several junctions with
different areas; the vertical error bars are derived from the
measured variation in $I_c R_N$ and a small noise contribution due
to the current source. From X-ray reflectivity results, as discussed
above, we take the error in $d_F$ for all F barriers to be $\approx
0.2$ nm. All of the devices shown in this data set presented Shapiro
steps upon the application of microwaves and an $I_c$ modulation
with applied field $H_A$. For the devices where $d_{Co}>5$ nm,
$d_{Py}>5$ nm, and $d_{Ni}>13$ nm some reduction of the differential
resistance around $I=0$ was seen, but did not show a measurable
supercurrent at $4.2$ K. As expected, $I_c R_N$ for the Co, Py and
Ni decreases exponentially and in an oscillatory fashion with $d_F$.

\begin{figure}[h]
\centering
\includegraphics[width=8cm]{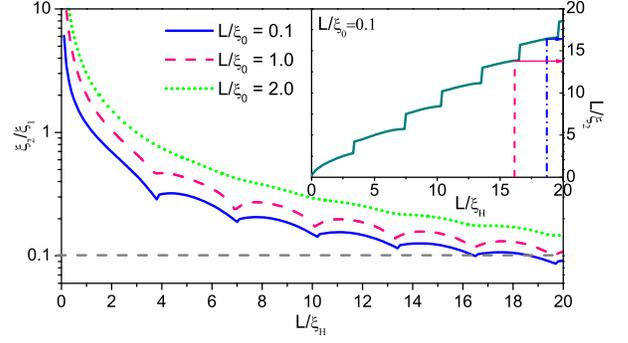}
\caption{(Color online) The dependence of $\xi_2 /\xi_1$ with
inverse magnetic length, $L/\xi_H$, calculated for different ratios
of $L/\xi_0$. Inset: inverse decay length, $L/\xi_2=f(L/\xi_0)$ for
when $L/\xi_H \simeq 0.1$ \label{paperBorn}}
\end{figure}

The Co data was modeled using Eq. (\ref{cleanlimit1}); from the
model fit shown in Fig. \ref{merge}(a) we find that the period of
the Co oscillations is $\simeq 1.9$ nm, hence $\xi_2 \simeq 0.3$ nm
and $\xi_1 \simeq 3.0$ nm. This gives a $\xi_2/\xi_1$ ratio of
$\sim0.11$. A full theoretical treatment involves solving linear
Eilenberger equations \cite{Gusakova} and gives Eq. (\ref{Born})
\begin{equation}
\ \tanh \frac{L}{\xi_{eff}}=\frac{\xi_{eff}^{-1}}{\xi_0^{-1} +
L^{-1} + i \xi_H^{-1}} \label{Born}
\end{equation}
where $\xi_{eff}$ is the effective decay length given by
$\xi_{eff}^{-1} = \xi_1^{-1} + i \xi_2^{-1}$, $\xi_o$ is the
Ginzburg-Landau coherence length and $\xi_H$ is a complex coherence
length. In the clean limit $1+L \xi_0^{-1}>>\frac{1}{2}$max$\{ \ln
(1 + L \xi_0^{-1}), \ln(L \xi_H^{-1})\}$. The solution of Eq.
(\ref{Born}) gives
\begin{equation}
\ \xi^{-1}_1 = \xi_0^{-1} + L^{-1}, \xi_0=\frac{v_F \hbar}{2 \pi T_c
k_B}, \xi_2 = \xi_H = \frac{v_F \hbar}{2 E_{ex}}, \label{Eilen}
\end{equation}
and the numerical solution is shown in Fig. (\ref{paperBorn}).

Following the method of Gusakova and Kupriyanov \cite{Gusakova} we
find from Fig. (\ref{paperBorn}) that the experimental ratio
$\xi_2/\xi_1\simeq 0.1$ corresponds to two inverse magnetic lengths
of $L/\xi_H\simeq 16.5$ and $L/\xi_H\simeq 18.7$. By assuming
$L/\xi_0\simeq 0.1$ and for the estimated parameters $\xi_1\simeq 3$
nm and $\xi_2\simeq 0.3$ nm we obtained, from the inset in Fig.
(\ref{paperBorn}) that for $L/\xi_H\simeq 16.5$ and $L/\xi_H\simeq
18.7$, $\overline{L}\simeq 5$ nm. Inputting these values into Eq.
(\ref{Eilen}) gives $v_F(Co)\simeq 2.8\times 10^5$ ms$^{-1}$ which
is similar to other reported values of $v_F(Co)$ \cite{Petrovykh98}
and $E_{ex}\simeq 309$ meV. To validate the mean free path of our Co
thin film we have estimated $L(Co)$ in a $50$ nm thick Co film by
measuring its resistivity as a function of temperature using the Van
der Pauw technique \cite{Gurney}. Following the method described by
Gurney \emph{et al.} we estimate that our Co has a mean free path of
$L(Co)\simeq 10$ nm which justifies our assumption of a clean limit
fit. As a comparison we have also modeled the Co oscillations with a
simpler theoretical model given by Eq. (\ref{cleanlimit})
\cite{BuzdinJ82}
\begin{equation}
\ I_c R_N \propto  \frac{\mid \sin(2 E_{ex} d_F /\hbar v_f) \mid}{2
E_{ex} d_F /\hbar v_f}. \label{cleanlimit}
\end{equation}
As in the case of Eq. (\ref{cleanlimit1}) the fitting between the
theoretical model and the experimental data is good (see Fig.
\ref{merge} (a) dashed line) and in particular the best fitting has
been obtained by using  $v_F=2.8\times 10^5$ ms$^{-1}$ and
$E_{ex}=309$ meV.

In contrast, the Py and Ni data cannot be fitted entirely in the
clean limit. In the case of Py, Eq. (\ref{cleanlimit}), closely
matches the experimental data up to a thickness of $\simeq 3.6$ nm
and in the case of Ni the oscillations follow the clean limit theory
to $\simeq 7$ nm. Above these values a better fit is obtained using
a formula for a diffusive and high $E_{ex}$ F \cite{Bergeret}:
\begin{equation}
\ I_C R_N \propto \mid Re \sum_{\omega_m >0}
\frac{\Delta^2}{\Delta^2 + \omega_m^2} \int^1_{-1}
\frac{\mu}{\sinh(k_\omega d_F / \mu L)}d\mu \mid, \label{dirtylimit}
\end{equation}

where $\Delta$ is the superconducting energy gap, $\omega_m$ is the
Matsubara frequency and is given by $\omega_m = \pi T k_B (2m+1)$
where $T$ is the transmission coefficient and $m$ is an integer
number. $k_\omega=(1+2\mid \omega_m \mid \tau / \hbar) - 2iE_{ex}
\tau/\hbar$ and $\mu=\cos \theta$ where $\theta$ is the angle the
momentum vector makes relative to the distance normal to the SF
interface. $L$ is given by $v_f \tau$ and $\tau$ is the momentum
relaxation time. For Eq. (\ref{cleanlimit}) the only fitting factor,
besides the numerical prefactor, was the strength of the exchange
interaction ($E_{ex}/\hbar v_f$). In the case of Eq.
(\ref{dirtylimit}) a suitable $v_f$, $\Delta$ and $E_{ex}$ had to be
chosen. To fit Eq. (\ref{dirtylimit}) to Py and Ni data we used:
$v_f (Py) = 2.2\times 10^5$ m/s and $L_{Py}\simeq 2.3$ nm, and $v_f
(Ni) = 2.8\times 10^5$ m/s and $L_{Ni}\simeq 7$ nm and $\Delta =
1.3$ meV. These values are consistent with the ones used in Eq.
(\ref{cleanlimit}) and elsewhere \cite{Blum,cbellprb05}. From the
oscillations period we can estimate the $E_{ex}$ of the Py and Ni
barriers - from Eq. (\ref{cleanlimit}) the periodicity is given by
$L_{osc}\sim \hbar v_f /2E_{ex}$ and hence we can deduce $E_{ex}$.
The exchange energies for Py and Ni were found to be $201$ meV and
$80$ meV respectively. $E_{ex}(Py)$ is approximately double that
measured in Nb/Py/Nb junctions deposited with epitaxial barriers
where $E_{ex}$ was found to be $95$ meV \cite{cbellprb05}.
$E_{ex}(Ni)$ is close to other reported values by photoemission
experiments \cite{Heinmann}. The smaller than expected $E_{ex}$ is a
consequence of impurities and possibly interdiffusion of Ni into Nb.
In the case of Ni, we have measured the magnetisation saturation as
a function of temperature ($M(T)$) (see inset of Fig.
\ref{merge}(c)) so that an estimate of the Ni Curie temperature
could be made. This provides further evidence that our Ni data is
consistent and that the Ni isn't grossly degraded.  To be sure our
data was not influenced by thermal diffusion during this measurement
we measured $M(T)$ for both the heating and cooling phases and it is
shown that the cooling curve follows the warming one. The warming
data is modeled by the formula: $M(T)/M(0)=(1-T/T_{curie})^{\beta}$
where $M(0)$ is the magnetisation at absolute 0 K, $T$ is the
measuring temperature, and $\beta$ and $T_{Curie}$ are fitting
parameters. We estimate $T_{curie}\simeq 571$ K which is in
agreement with Curie temperature measurements of Ni in S/F bilayers
\cite{Kim2005}.

\begin{table}
\begin{tabular}{l c c c c c c} \hline
$\xi_1$ (nm) & $\xi_2$ (nm) & F & $v_F$ (ms$^{-1}$) & $E_{ex}$ (meV) & DL(nm) & Ref \\
\hline \hline
1.2 & 1.6 & Ni$_{20}$Fe$_{80}$ & 2.2 $\times 10^5$& 95 & 0.7 &\cite{cbellprb05}\\
1.4 & 0.46 & Ni$_{20}$Fe$_{80}$ & 2.2 $\times 10^5$ & 201 & 0.5 & *\\
1.8 & 2.0 & Pd$_{90}$Ni$_{10}$ & 2.0 $\times 10^5$ & 35 & - &\cite{Kontos02}\\
1.7 & 1.0 & Ni & 2.8 $\times 10^5$ & 200 & - & \cite{Blum2006} \\
2.3 & 0.86 & Ni & 2.8 $\times 10^5$ & 107 & - & \cite{Blum} \\
4.1 & 1.2 & Ni & 2.8 $\times 10^5$ & 80 & 1.5 & *\\
4.6 & 0.45 & Ni$_3$Al & 1.5 $\times 10^5$& 86 & 5-8 & \cite{Born2006}\\
3.0 & 0.3 & Co & 2.8 $\times 10^5$ & 309 & 0.8 & *\\
\hline
\end{tabular}
\caption{A summary of reported parameters for different material
systems. DL stands for `Dead Layer Thickness' and * `This
paper'.\label{table}}
\end{table}


The fully clean limit behavior observed with the Co barriers arises
most obviously from the use of a pure element, but also from the
vertical coherence likely \cite{Leung} even in noncrystalline
heterostructures. The high $E_{ex}$ results in a short oscillation
period implying a need for \AA-level control of layer thickness for
practical devices; however Co and Co-alloys form the basis of
current spintronic device production, and precision layer control
and excellent compatibility with tunnel barriers \cite{Oleinik} have
been demonstrated in many industrial processes \cite{Kaisera}. Co is
an attractive material for qubits and other novel devices.

In summary, we have measured critical current oscillations in Co
junctions as a function of Co barrier thickness which indicates that
the devices are strongly in the clean limit. This results in a
higher $I_c R_N$ values in the $\pi$ state compared to $I_c R_N$
values in the dirty limit. We also present complementary critical
current oscillations through Py and Ni barriers. The oscillations in
$I_c R_N$ with $d_F$ are indicative of $0-\pi$ crossovers, but
without conducting phase-sensitive measurements the specific phases
of the individual junctions cannot be known. The oscillations also
showed an excellent fit to theoretical models. We have also
estimated, from the periodicity of the oscillations, the exchange
energies of the Co, Py and Ni barriers to be $309$ meV, $201$ meV
and $80$ meV respectively. Results within this paper are summarized
in table \ref{table} alongside results reported elsewhere. Our
results are not only interesting in their own right, but are a vital
experimental step towards understanding the physics of quantum
electronic devices based on superconductors and are of considerable
value to the development of quantum information processing. Our
devices are precursors to practical implementations into qubits and
other applications in controllable and scalable superconducting
quantum electronic devices.

The authors thank M. Yu. Kupriyanov, A. M. Cucolo, and F. Bobba for
helpful discussions and N. Stelmashenko for technical assistance. We
acknowledge the support of the EPSRC UK and the European Science
Foundation $\pi$-shift network.


\end{document}